# Association Rules Mining Based Clinical Observations

**Mahmood A. Rashid[1], Md Tamjidul Hoque[2], Abdul Sattar[1]**

[1]*Institute for Integrated and Intelligent Systems (IIIS),*
[2]*Discovery Biology, Eskitis Institute for Cell & Molecular Therapies,*
*Griffith University*
*Nathan, QLD, Australia*
*{m.rashid, t.hoque, a.sattar}@griffith.edu.au*

*Abstract*

*Healthcare institutes enrich the repository of patients' disease related information in an increasing manner which could have been more useful by carrying out relational analysis. Data mining algorithms are proven to be quite useful in exploring useful correlations from larger data repositories. In this paper we have implemented Association Rules mining based a novel idea for finding co-occurrences of diseases carried by a patient using the healthcare repository. We have developed a system-prototype for Clinical State Correlation Prediction (CSCP) which extracts data from patients' healthcare database, transforms the OLTP data into a Data Warehouse by generating association rules. The CSCP system helps reveal relations among the diseases. The CSCP system predicts the correlation(s) among primary disease (the disease for which the patient visits the doctor) and secondary disease/s (which is/are other associated disease/s carried by the same patient having the primary disease).*

*Key words*

*Disease Correlation, Association Mining, e-Health, Healthcare, Medicare*

## 1. Introduction

In the recent era, medical science has revealed that the occurrence of one disease can lead to several associated diseases [1]. For example, Heart-Block can lead to the occurrences of other diseases like Hypertension, Cardiac-Arrest and so on. It is, however, still an interesting problem [1], to see how far the medical philosophy holds from statistical point of view. Data mining based techniques, like association rule mining, have gained popularity [1-3] among contemporary scientists to gain clearer understanding of different physical and scientific phenomenon. In this paper, we apply association rule mining to extract knowledge from clinical data for predicting correlation of diseases carried by a patient.

From the viewpoint of scientific research, data mining is relatively a new discipline that has been developed mainly from studies carried out in various disciplines such as computing, marketing, statistics and so on [4]. Data mining problems and corresponding solutions have roots in classical data analysis. Many of the methodologies used in data mining has come from two branches of research: *i)* one is developed in machine leaning (artificial intelligence) community, and *ii)* the other is developed in the statistical community particularly in multivariate and computational statistics. Both have made great contributions to the understanding and applications of data mining techniques [5, 6].

The remainder of the paper is organized as follows. Section 2 highlights previous research in related areas. Section 3 introduces the Architecture of the CSCP Prototype-system. It also briefly discusses the working procedures and Algorithms used to compute the correlations. Section 4 highlights the output of the system with the hypothetical datasets. Concluding remarks and future research are sketched in section 5.

## 2. Related Works

Automated healthcare systems are accumulating large quantities of information about patients and their medical conditions everyday. Unfortunately, few methodologies have been developed and applied to discover this hidden knowledge [7].

The cluster-analysis based model is suggested and discussed [8] for assigning prostate cancer patients into homogenous groups with the aim to support future clinical treatment decisions as an illustration. To explore association rules in noisy and high dimensional medical data-repository an improved algorithm has been introduced with several constraints [9]. A statistical analysis of decision tree based classification approach on diagnosing the Ovarian Cancer using Bio-marker Patterns Software (BPS) has been applied [10]. A task [11] has been accomplished on comparison of data mining methods supporting diagnosis for Melanoma. Association rule classifiers have been applied to diagnose breast cancer using digital mammograms [12], Neural Network based classification approach also used for the same purpose [13]. Association Mining applied on questionnaire responses related to human sleeping [14] where questionnaire data and clinical summaries comprised a total of 63 variables including gender, age, body mass index, and Epworth and depression scores.

Many Clinical Decision Support Systems (CDSS) have been developed. CHICA [15] is a CDSS, developed to improve preventive paediatric primary care. Dynamic forms are generated and tailored to patients' needs based on the Medical Logic Modules (MLMs). A knowledge management framework for distributed healthcare systems has been proposed [16] to integrate the heterogeneous systems used by different departments from clinical care to administration.

However, the aforementioned developments are application specific and thus hard to apply in general. Instead of developing an application limited to a specific purpose such as prostate cancer [8], skin cancer [11], and sleeping [14] and so on, we proposed for a more generic version of CSCP system that can work for all diseases in similar fashion and generate correlations depending on the input dataset. In the next section, the architecture of the application and its working procedures are stated.

## 3. Framework and Working Procedures

The CSCP system will extract data from an OLTP system. Thus to implement the CSCP system, we require the regular operational OLTP system to generate input data for the CSCP system. The architecture of the system can be described as follows.

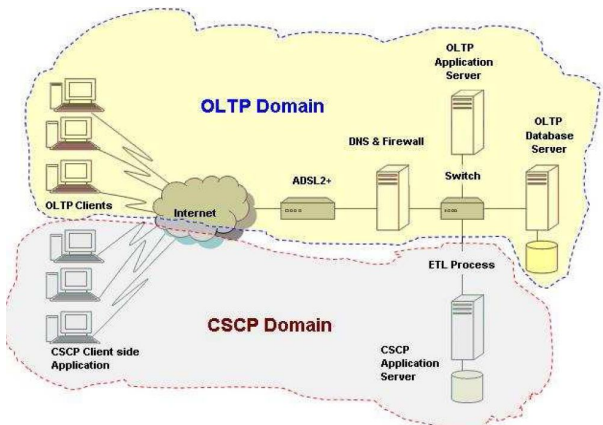

*Figure 1 - The Complete Architecture of the Proposed System for Prediction Correlation among diseases using association mining on healthcare data. The upper-portion of the figure describes the architecture of the OLTP system and the lower portion is the CSCP system.*

### 3.1 System Architecture

The disease diagnosis application proposed here based on two different software systems. One is online transaction processing (OLTP) system and another is disease correlation predicting (CSCP) application.

*OLTP Application:* When a patient will visit a doctor he/she must have to fill up a prescribed form and the information from the form is proposed to be captured through web enabled OLTP system.

*CSCP Application:* This application will import transactional records from OLTP application for further processing to generate correlations among diseases using Association Rules data mining.

The OLTP system is a regular database application to capture patient's information and to preserve the records into a database repository. It is simple and no analytical process has been incorporated in this portion and thus the CSCP system is our major focus in this paper and we presume that the OLTP system has been implemented successfully and data has been captured accordingly from healthcare institutes.

### 3.2 Association Rules Mining and Apriori Algorithm

In data mining, association rule is used for discovering interesting relations between variables in large databases. The two key terms support and confidence are used in computing correlations between variables which can be defined as follows:

*Support:* In a fixed number of transactions the occurrences of a particular event is the support of that event. For example there are T transactions among which $T_{xy}$ transaction s contain the itemset {X,Y} is $T_x$.

*Confidence:* confidence is a relative support. For example there are T transactions among which $T_{xy}$ transactions contain itemset {X, Y} and $T_x$ transactions contain item X and thus confidence of occurring X and Y together is $T_{xy}/T_x$.

The Apriori Algorithm [17] is used here for Association Rule Mining to find out frequent dataset that satisfy the predefined minimum support and confidence from a given database. As computers are handy now-a-days, a viable CSCP system can be setup to collect large volumes of data and to stored in the database simultaneously. This kind of data includes the transaction records of clinics, hospitals, supermarkets, banks, stock markets, telephone companies and so on. The next few sections we try to discover some hidden information from a sample transactional dataset.

### 3.3 Sample Dataset

In a clinic various patients come but most of them come for a particular disease, which mentioned here as primary disease. When Doctors or their associates make an interview with the patient and note down other problems (diseases), which mentioned here as secondary or associated diseases are inserted into a database.

The main objective of the paper is to find out relations among the primary disease and other secondary diseases. The following table represents sample dataset of a Medicare database that contains the patient-wise diseases.

*Table 1 - Transactional data sample*

| Patient Id | Disease |
|---|---|
| P000000001 | Bradycardia |
| P000000001 | Cardiac Arrest |
| P000000001 | Hypertension |
| P000000001 | Myocarditis |
| P000000002 | Bradycardia |
| P000000002 | Cardiac Arrest |
| P000000002 | Hypertension |
| … | … |
| P000001000 | Cardiac Arrest |

### 3.4 Producing Itemsets

A set of diseases obtained by each patient presented here along with the number of diseases counted by the algorithm in Figure 3 and put the result into the column headed CNT in figure 3

*Algorithm 1 - CountDisease*

| | |
|---|---|
| 1 | PROCEDURE CountDisease |
| 2 | FOR each p in P |
| 3 | $D_s \Leftarrow ""$ |
| 4 | $c \Leftarrow 0$ |
| 5 | find records for p |
| 6 | FOR each r in R |
| 7 | $c \Leftarrow c+1$ |
| 8 | $D_s \Leftarrow D_s + r + ","$ |
| 9 | NEXT r |
| 10 | $D_s \Leftarrow D_s$ without comma at the end |
| 11 | INSERT $(p,c,D_s)$ in to database table |
| 12 | NEXT p |
| 13 | END CountDisease |

*Figure 2 - Pseudo code to count diseases carried by any patient.*

*Table 2: Patient records with multiple diseases*

| Patient | Count | Diseases |
|---|---|---|
| P000000001 | 4 | Heart-Block, Hypertension, Cardiac-Arrest, Bradycardia |
| P000000002 | 3 | Heart-Block, Hypertension, Cardiac-Arrest |
| … | … | … |
| P000001000 | 1 | Hypertension |

### 3.5 Counting Support of an Item (Disease) in the sample dataset

The frequency of every item in all the transactions has been calculated in the following table implementing the following algorithm and for a 1-item itemsets for the first pass.

*Algorithm 2- FindSupport*

| | |
|---|---|
| 1 | PROCEDURE FindSupport |
| 2 | $d \Leftarrow disease$ |
| 3 | db $\Leftarrow$ database in consideration |
| 4 | $r \Leftarrow record$ |
| 5 | $rs \Leftarrow recordset$ |
| 6 | $s \Leftarrow support$ |
| 7 | BEGIN |
| 8 | FOR each d in db |
| 9 | $s \Leftarrow 0$ |
| 10 | find records for d |
| 11 | FOR each r in rs |
| 12 | $s \Leftarrow s+1$ |
| 13 | NEXT r |
| 14 | UPDATE Database with s |
| 15 | NEXT d |
| 16 | END |
| 17 | END FindSupport |

*Figure 3 - Pseudo code to calculate support.*

*Table 3 - Distinct diseases*

| Disease | Support | Pass |
|---|---|---|
| Heart-Block | 334 | 1 |
| Hypertension | 549 | 1 |
| Myocarditis | 532 | 1 |
| Cardiac-Arrest | 536 | 1 |
| Bradycardia | 305 | 1 |

### 3.6 Generating Candidate Itemset

The following procedure generates candidate itemsets taking the transactional records as input and maximum pass and minimum support as parameters.

*Algorithm 3 - GenerateRule*

PROCEDURE GenerateRule (maxpass, minsup)
    // the method requires two user input
    // maxpass- maximum number of items are consider in an itemset
    // minsup- minimum support for the candidate sets
BEGIN
1. Produce *maxpass* copies of transaction data by aliasing the original one for generating all possible combination of items in each pass by producing Cartesian product and the filter the meaning sets
2. Calculate the supports for the filtered sets
3. Eliminate the sets having support less than minsup

END

*Figure 4 - Generating candidate itemset from sample dataset.*

*Table 4 - Candidate Sets for stop-level 2*

| Itemset | Count |
|---|---|
| {Bradycardia, Cardiac-Arrest} | 26 |
| {Bradycardia, Heart Block} | 9 |
| {Bradycardia, Hypertension} | 28 |
| {Bradycardia, Myocarditis} | 21 |
| {Cardiac Arrest, Heart-Block} | 13 |
| {Cardiac Arrest, Myocarditis} | 32 |
| {Heart Block, Hypertension} | 10 |
| {Heart Block, Myocarditis} | 11 |

### 3.7 Association Rule Generation

Using the sample transactions of Table 1, after second pass (maxpass=2), the CSCP system has generated the rule data as listed in Table 5 below

*Table 5- Candidate Sets for stop-level 2*

| Itemset | Supp (%) | Conf (%) |
|---|---|---|
| {Bradycardia, Cardiac-Arrest} | 2.60 | 8.52 |
| {Bradycardia, Heart-Block} | 0.90 | 2.69 |
| ………… | ….. | ….. |
| {Heart Block, Hypertension} | 1.00 | 1.82 |
| {Heart Block, Myocarditis} | 1.10 | 3.29 |

## 4. Discussion on the Empirical Results

In practical applications, a rule can be considered statistically significant when it generates from a data repository that contain hundred of thousands or millions of trusted transactions. We have carried out our experiments on a relatively small number of transactions[1] due to the lack of availability of real transactional data where the disease Heart-Block and the disease Hypertension conjunctionally occurs once and thus its support of {Heart Block , Hypertension} is 1.0 as shown in *Table 5*.

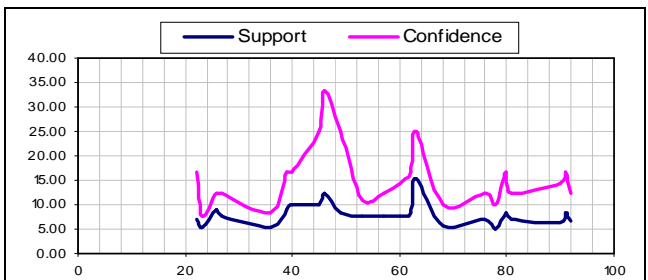

*Figure 5- For the itemset {Hypertension, Heart-Block} in second pass, the support and confidence for patients of different age groups. X-axis represents the ages and Y-axis represents the value of support and confidence respectively.*

---
[1] Randomly generated 1000 transactions

*Figure 5* represents the support and confidence of second pass for the disease set{Hypertension, Heart-Block} for the patients of different age groups.

The graph in the *Figure 6* represents the supports and confidences for different itemsets for the female patients in the sample transactions.

In our work we implement the system to compute the correlation of diseases for different age groups and the output in *Figure 5* is clearly illustrating that at the age of 45-50 the risk of Heart-Block is more than other ages if a patient has already been living with Hypertension. Our system has also been applied on the database for the patent of different sexualities and the graphical form of empirical output has been shown in *Figure 6* which illustrates that for the female patients the risk of Hypertension is significant if the patient already carry Bradycardia.

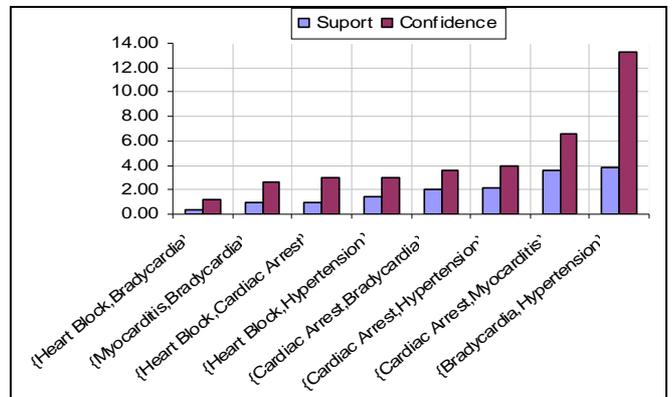

*Figure 6 - For the different itemsets in second pass, the supports and confidences for patients of different sexualities (here females only). X-axis represents the disease sets and Y-axis represents the value of support and confidence respectively.*

The system can easily be enhanced for further crucial aspects: such as patients of various professions, localities, physiological conditions and so on. The nature of working environment for IT professionals is remarkably different from that of the construction professionals. IT professionals work in static indoor environment where as construction professionals work in dynamic and outdoor environment. By implementing our system on the patients' database of IT profession and construction profession, new trend of disease correlation can be predicted.

## 5. Conclusions

In this paper, we have implemented system-prototype, named CSCP system, using the association rules of data mining technique applied to a patients' (assumed) database for discovering patterns of diseases that might be carried by a patients. As a novel idea of mining the data capturing process can further be modified in the clinics as well as in the data-warehouses which should further be involved to enhance the CSCP system we have proposed. The recognised pattern by this implementation definitely can improve the healthcare services along with medical researchers for further exploring trends of diseases that are correlated. To ensure strong national economy and bio-security [18] by having healthier inhabitants, for example, Medicare Australia [19] can use CSCP system to ensure wellbeing further.

In this research-work, we have succeeded to investigate correlation among diseases for patients of different age and sex groups providing the outcome in statistical as well as in graphical format.

The system we developed has been based on computer generated data, since the real data were not handy. However, it is our future target to enhance the system for the aforementioned various cases, applying the system on collected real-life data while enhancing the proposed system rigorously.

## 6. Acknowledgement

This research is partially supported by the ARC (Australian Research Council) grant no DP0557303.

**Address for correspondence**

*Mahmood A. Rashid*
Building N34, Room 1.45
Institute for Integrated and Intelligent Systems (IIIS)
Griffith University, Nathan, QLD 4111, Australia
Email: mahmood.rashid@gmail.com
Phone: +61(0)7 373 53757, Fax: +61 (0)7 373 54066